\documentclass[twocolumn,showpacs,aps,prl]{revtex4}

\newcommand{\bk}{{\bf k}}

\newcommand{\beq}{\begin{eqnarray}}
\newcommand{\eeq}{\end{eqnarray}}
\newcommand{\beqq}{\begin{eqnarray*}}
\newcommand{\eeqq}{\end{eqnarray*}}

\usepackage{graphicx}
\usepackage{dcolumn}
\usepackage{bm}
\usepackage{color}

\begin{document}

\begin{titlepage}

\title{Surface States of Topological Crystalline Insulators in IV-VI Semiconductors}

\author{Junwei Liu$^{1,2}$, Wenhui Duan$^1$ and Liang Fu$^2$}
\address{$^1$Department of Physics and State Key Laboratory of Low-Dimensional Quantum Physics, Tsinghua University, Beijing 100084, People's Republic of China\\ $^2$Department of Physics, Massachusetts Institute of Technology, Cambridge, MA 02139}

\begin{abstract}
Topological crystalline insulators (TCI) are new topological phases of matter protected by crystal symmetry of solids. 
Recently, the first realization of TCI has been predicted and observed in IV-VI semiconductor  SnTe and related alloys Pb$_{1-x}$Sn$_x$(Te, Se). 
By combining $k \cdot p$ theory and band structure calculation, we present a unified approach to study topological surface states on various crystal surfaces of TCI in IV-VI semiconductors. We explicitly derive $k \cdot p$ Hamiltonian for topological surface states from electronic structure of the bulk, thereby providing a microscopic understanding of bulk-boundary correspondence in TCI. 
Depending on the surface orientation, we find  two types of surface states with qualitatively different properties. In particular, we predict that (111) surface states consist of four Dirac cones centered at time-reversal-invariant momenta $\bar \Gamma$ and $\bar M$, while (110) surface states consist of Dirac cones at non-time-reversal-invariant momenta, similar to (001). Moreover, both (001) and (110) surface states exhibit a Lifshitz transition as a function of Fermi energy, which is accompanied by a Van-Hove singularity in density of states arising from saddle points in the band structure.   
\end{abstract}


\maketitle

\draft

\vspace{2mm}

\end{titlepage}

Structure and symmetry play an important role in shaping electronic properties of periodic
solids. It is a common phenomenon that materials made of chemically similar elements 
arranged in the same crystal structure, like diamond and silicon, often have qualitatively similar electronic properties. This empirical  
relationship between 
structure and property arises from the fact that the essential electronic properties of many solids are  understandable in terms of 
orbitals and bonds, the characteristics of which depend mostly on crystal structure. For example, both diamond and silicon 
possess $sp^3$ hybridized orbitals in tetragonal structure.  
On the other hand, the quantum theory of solids is based on itinerant Bloch waves that form energy bands in momentum space. 
The global structure of band theory allows for unconventional energy bands 
that are topologically nontrivial as a whole entity.  
Such band structures are fundamentally different 
from a Slater determinant of atomic orbitals, and give rise to topological states of matter exhibiting universal and 
quantized properties that are absent in ordinary solids.  Celebrated examples of such topological states include 
quantum Hall insulators\cite{haldane} and topological insulators\cite{kane, zhang, moore}.

The topological aspect of band theory gives birth to interesting exceptions to the empirical rule of structure-property relation. 
We  demonstrated\cite{fu}  
the proof of principle that  there exist distinct classes of band structures  within the same crystal structure, which  
cannot be adiabatically connected under deformations preserving certain point group symmetries. 
Those nontrivial band structures are characterized by topological indices, and thereby define
a new topological state of matter dubbed 
topological crystalline insulators (TCI). A hallmark of TCI is the existence of gapless surface states on those surfaces that 
preserve the underlying crystal symmetry\cite{mong}. 

Recently, Hsieh {\it et al} predicted\cite{hsieh} the first material realization 
of TCI in IV-VI semiconductors SnTe and related alloys Pb$_x$Sn$_{1-x}$(Te, Se).  
The nontrivial band topology in these materials is characterized by an integer topological invariant known as mirror Chern number\cite{teofukane}, 
arising from  the mirror symmetry with respect to the $(110)$ plane of 
the rocksalt structure (and its symmetry-related ones). A consequence of this electronic topology is the existence of 
topological surface states on crystal surfaces that preserve at least one such mirror symmetry, such as (001), (111), (110).  
In particular, the (001) surface states consist of four Dirac cones at low energy, which are located 
at generic point $\bar \Lambda$ on the mirror-symmetric line $\bar{\Gamma} \bar X$. 
Importantly, these four Dirac cones are spin-momentum locked with 
the {\it same} chirality, which is a unique hallmark of the TCI phase in IV-VI semiconductors. 
These predicted surface states were subsequently observed in 
 angle-resolved photoemission spectroscopy experiments on 
SnTe\cite{ando}, Pb$_{1-x}$Sn$_{x}$Se\cite{poland} and Pb$_x$Sn$_{1-x}$Te\cite{hasan}. 
In particular, the spin-texture of these surface states observed by Xu {\it et al}\cite{hasan} (see also Ref.\cite{spin}) provides a direct spectroscopic measurement of the mirror Chern number\cite{sb, murakami}.   

The materialization of TCI opens up a new venue for topological states of matter in a much larger number of material classes than 
previously thought\cite{news, fiete, fang, teohughes, yao}, thereby triggering  
intensive activities\cite{fukane, anton, pbte, xiao, zaanen}. From a material viewpoint, IV-VI semiconductors have high mobilities 
and exhibit a wide range of electronic properties (e.g.,    
magnetism, ferroelectricity and superconductivity) that can be easily tuned by alloying, doping and strain. 
The technology for synthesizing and engineering these materials, in both bulk and low-dimensional form, has been well-developed by decades of efforts.   
Therefore TCI in IV-VI semiconductors provide an extremely versatile platform for exploring topological quantum phenomena and novel device applications.

In this work, we combine $k \cdot p$ theory and band structure calculation to study topological surface states on various crystal surfaces of TCI in IV-VI semiconductors. We present a unified approach to derive $k\cdot p$ Hamiltonian for surface states from the electronic structure of the bulk, thereby providing a {\it microscopic} understanding of bulk-boundary correspondence in TCI. 
We find that low-energy properties of surface states are solely determined by the surface orientation, and can be classified into two types.  
 In particular, we predict that (111) surface states consist of four Dirac cones centered at time-reversal-invariant momenta $\bar \Gamma$ and $\bar M$\cite{aps}, while (110) surface states consist of Dirac cones at non-time-reversal-invariant momenta, similar to (001). Moreover, both (001) and (110) surface states exhibit a Lifshitz transition  as a function of  Fermi energy, which is accompanied by a Van-Hove singularity in density of states arising from saddle points in the band structure. Our results provide a much-needed basis for further investigations on  TCI surface states. 

We begin by reviewing $k\cdot p$ theory for the bulk band structure of TCI, from which surface states are derived.  
The band gap of IV-VI semiconductors is located at four $L$ points. 
For the ionic insulator PbTe, the Bloch state of the valence band at $L$   
is derived from the $p$-orbitals of the anion Te, and that of the conduction band  
from the cation Pb. In contrast, SnTe 
has an inherently inverted band ordering, in which the valence band is derived from the cation Sn 
and the conduction band from Te.  
This band inversion relative to a trivial ionic insulator gives rise to the TCI phase in SnTe\cite{hsieh}. 

 The band structure near each $L$ point can be described by $k\cdot p$ theory  
 in the basis of the four Bloch states at $L$, $\psi_{\sigma, s}(L)$, where $\sigma=1(-1)$ refers to the state derived from 
the cation (anion), and $s$ labels the Kramers degeneracy. 
The $k \cdot p$ Hamiltonian $H(\bk)$ is given by (see \cite{hsieh} and references therein): 
\beq
H({\bf k})= m \sigma_z + v \sigma_x (k_1 s_2 - k_2 s_1) + v' k_3 \sigma_y,  
\label{kp}
\eeq  
where $k_3$ is along the $\Gamma L$ direction, and $k_1$ is along the $(1\bar{1}0)$ axis of reflection. 
$\vec \sigma$ 
and $\vec s$ are two sets of Pauli matrices.  
The sign of $m$ in (\ref{kp}) captures the two types of band ordering: in our convention $m>0$ for PbTe and $m<0$ for SnTe.   

\begin{figure}[tbp]
\includegraphics[width=8cm]{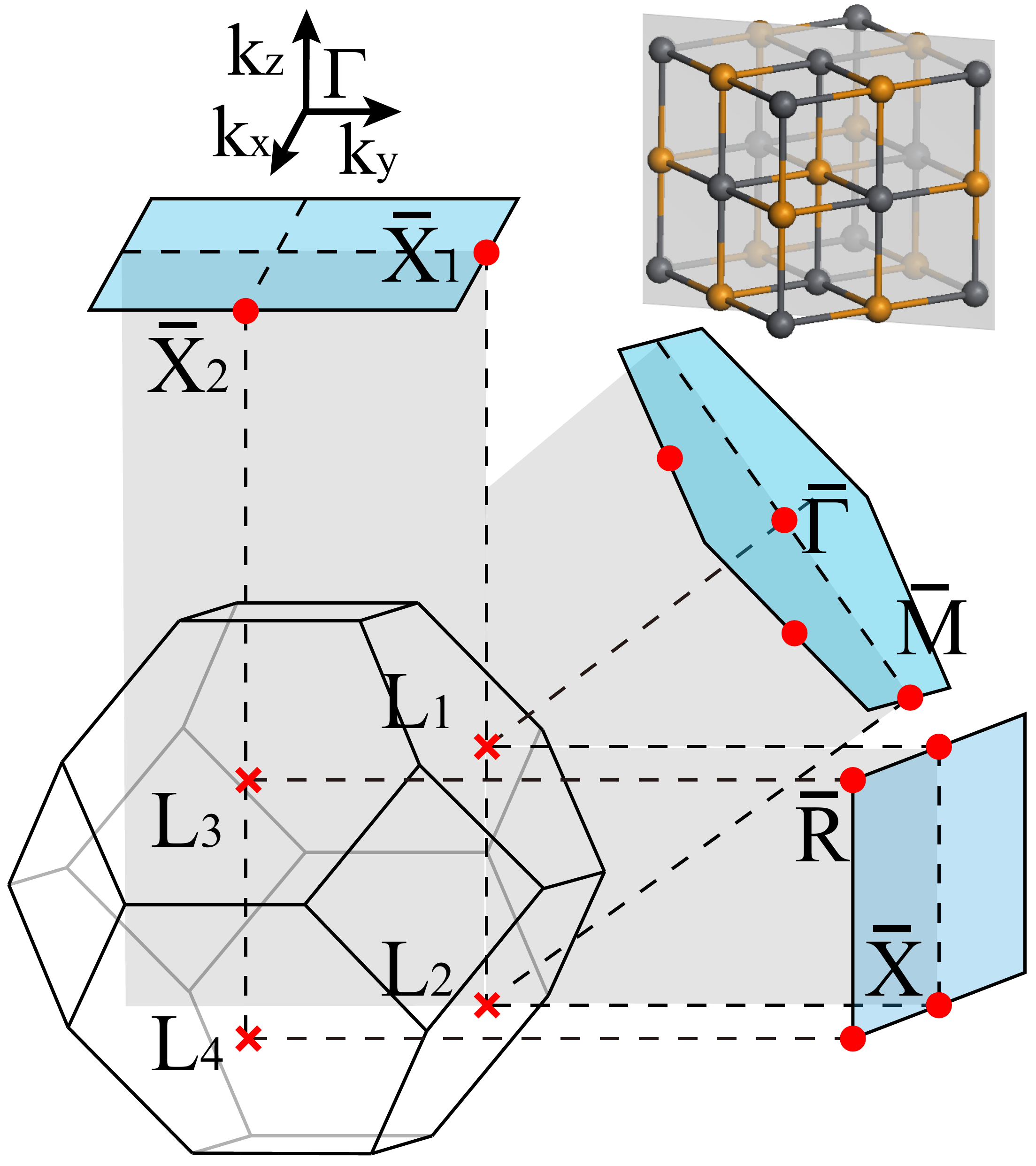}
\caption{Bulk Brillouin zone of rocksalt structure and its projection to (001), (111) and (110) surface Brillouin zone.  
For (001) surface, $L_1$ and $L_2$ are both projected to $\bar{X}_1$, and $L_3$ and $L_4$ are projected to $\bar{X}_2$; For (111) surface, $L_1$ is projected to $\bar \Gamma$ and the other three $L$ points are projected to $\bar M$ points;
For (110) direction, $L_1$ and $L_2$ are projected to $\bar{X}$ and $L_3$ and $L_4$ are projected to $\bar{R}$. 
The shaded plane passing through $\Gamma$, $L_1$ and $L_2$ in 3D Brillouin zone is invariant under reflection 
about $(1\bar10)$ plane in real space.
}
\end{figure}

Electronic structures of TCI surface states depend crucially on crystal face orientations. 
We find surface states with qualitatively different properties on two types of crystal surfaces. 
For type-I surface, all four $L$ points  are projected to different time-reversal-invariant momenta in the surface Brillouin zone. 
This is the case for (111) surface, for which $L_1$ is projected to $\bar \Gamma$ and $(L_2, L_3, L_4)$ are projected to 
three $\bar M$ points. For type-II surface, different $L$ points are projected to the same surface momenta. This is the case for  
(001) surface for which $(L_1, L_2) \rightarrow \bar X_1$ and $(L_3, L_4) \rightarrow \bar X_2$, 
as well as (110) surface for which $(L_1, L_2) \rightarrow \bar X$ and $(L_3, L_4) \rightarrow \bar R$.  
The projection from bulk to surface Brillouin zone is illustrated in Fig.1.

{\bf (111) surface}: 
type-I surface states 
can be obtained straight-forwardly from the continuum $k\cdot p$ Hamiltonian (\ref{kp}).  
Following the spirit of Ref.\cite{zkm}, we  
model vacuum as a trivial insulator (like PbTe) with an infinite gap $m=M>0$, where $M\rightarrow +\infty$. 
Surface states can now be obtained by solving a domain wall problem in which the Dirac mass $m$ changes sign across the interface between TCI and vacuum. It is well-known from field theory that  two-dimensional massless Dirac fermions form at such an interface\cite{volkov, drew, fradkin}. Due to the presence of four $L$-valleys, surface states consist of four copies of such Dirac fermions.  These four   Dirac fermions are located at four {\it distinct} 2D momenta that correspond 
to the projection of the four $L$ points onto the type-I surface. Because of their different in-plane momenta, 
the four $L$-valleys cannot couple with each other as long as  in-plane translation symmetry is present, and therefore  
independently give birth to four branches of Dirac surface states. 

A prime example of type-I surface is (111). It follows from the above analysis that (111) 
surface states consist of four Dirac cones: one at $\bar \Gamma$, and three others at $\bar M$. 
The $k\cdot p$ Hamiltonians 
at $\bar \Gamma$ and $\bar M$ are given by 
\beq
H_{\bar \Gamma} (\bk) &=& v (k_1 s_2  - k_2 s_1) \nonumber \\
H_{\bar M} (\bk) &=& v_1 k_1  s_2  - v_2  k_2 s_1 
\eeq
where $k_1$ is along $\bar \Gamma \bar K$ direction, and $k_2$ is along the mirror-invariant $\bar \Gamma \bar M$ direction. 
The presence of these Dirac pockets is confirmed by our band structure calculations on SnTe, based on the tight-binding model\cite{lent}, see Fig.2. The Dirac points are found to lie close to the top (bottom) of the valence (conduction) band for Sn (Te) termination. Such surface states are qualitatively similar to interface states between PbTe and SnTe studied in early theoretical works\cite{volkov, drew, fradkin}. The advent of TCI has revealed that such interface states are topologically equivalent to surface states of SnTe (but not PbTe), and their robust existence is topologically protected by the (110) mirror symmetry. 
This protection can be understood from the two branches of 
counter-propagating surface states on the mirror-symmetric  line $\bar{\Gamma} \bar{M}$.
Our tight-binding calculation shows that these two branches have opposite mirror eigenvalues and therefore cannot couple with each other to open up a gap, as long as mirror symmetry is preserved. This unusual property of (111) surface states 
agrees with the prediction based on  mirror Chern number\cite{hsieh}, and constitutes another important hallmark of TCI. 
Since mirror eigenvalue of a spin-1/2 fermion is intimately related to its spin state,  
it follows that spin-textures on all four Dirac pockets, at both $\bar \Gamma$ and $\bar M$,   
have the {\it same} chirality. 
 
\begin{figure}[tbp]
\includegraphics[width=8cm]{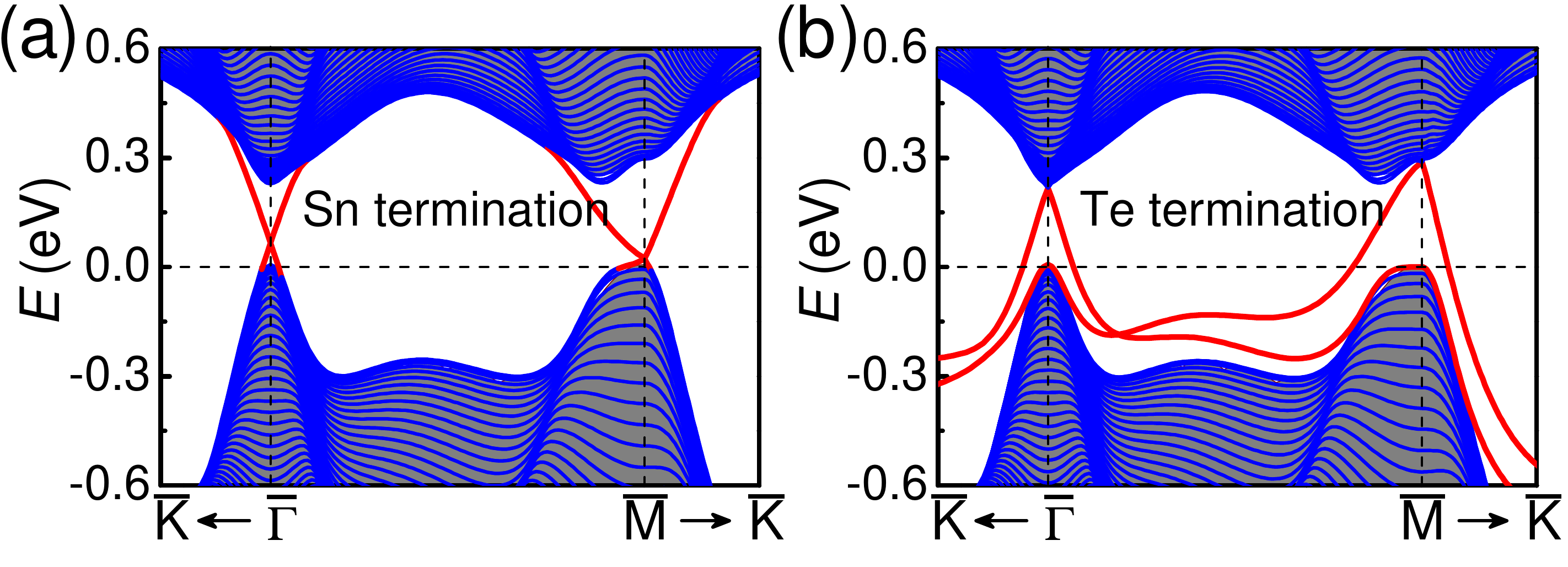}
\caption{Band structure of (111) surface for Sn and Te termination. Four Dirac pockets are present: one at $\bar \Gamma$ and three at $\bar M$. This leads to two counter-propagating states with opposite mirror eigenvalues on $\bar \Gamma \bar M$, as predicted from the mirror Chern number $n_M=-2$ in the TCI phase\cite{hsieh}. }
\end{figure}

{\bf (001) surface}:  type-II surfaces are much more interesting. 
In this case,  two $L$ points are projected to the same momentum in the surface Brillouin zone. As a result,  
they interact with each other to create topological surface states that are different from type-I surfaces. This interaction between $L$-valleys arises from physics at the lattice scale, which is not captured by the previous field-theoretic approach\cite{volkov, drew, fradkin}. For example, the previous ab-initio calculation\cite{hsieh} shows that (001) surface has a pair of Dirac cones {\it away from} $\bar X$ (the projection of $L$), which cannot be understood from field theory alone.  

To obtain type-II surface states, we proceed in two steps. First, we study 
a hypothetical smooth interface between TCI and a trivial insulator, in which the Dirac mass is slowly-varying in space and gradually changes from negative to positive over many lattice constants across the interface. Scattering between the two $L$ valleys projected to the same momentum on the surface requires large-momentum transfer in the direction normal to the interface and hence vanishes in this smooth limit. We are thus justified to treat different valleys independently and use the continuum field theory to solve for interface states. 
The effective Hamiltonian that we derive for these interface states serves as a starting point for surface states. 
Next, imagine deforming the smooth interface into the atomically sharp  surface, which adiabatically connects  interface states to the actual surface states. This deformation procedure 
introduces inter-valley scattering processes at the lattice scale, which are represented by additional terms in the effective Hamiltonian for the surface. Such terms must satisfy all the crystal symmetry at the surface, and therefore can be enumerated by a symmetry analysis. By incorporating these terms into the surface Hamiltonian derived in the previous step, 
we  obtain the final form of $k \cdot p$ theory for type-II surfaces.

We now apply this approach to study (001) surface. Starting from a smooth interface, there exist two  
massless Dirac fermions at ${\bar X}_1$ arising from the $L_1$ and $L_2$   
valley respectively, and likewise for the symmetry-related point $\bar X_2$. 
These two flavors of Dirac fermions 
have identical energy-momentum dispersions, resulting in a two-fold degeneracy at every $\bk$.  
The $k \cdot p$ Hamiltonian for this smooth interface  is given by
\beq
H_{{\bar X}_1}^0(\bk) =  (v_x k_x  s_y - v_y k_y s_x) \otimes I,  \label{h0}
\eeq
where the momentum $(k_x, k_y)$ is measured from $\bar{X}_1$, with $k_x$ parallel to $\bar \Gamma {\bar X}_2$ and $k_y$ parallel to $\bar \Gamma {\bar X}_1$. 
Here $I$ is identity operator in the flavor space, 
and $\vec s$ is a set of Pauli matrices associated with the two components of each flavor that are intimately related 
to electron's spin. 
The velocities in $x$ and $y$ directions are generically different.   

Next, we perform a symmetry analysis to deduce the form of those additional terms associated with physics at the lattice scale, which need to be  
added to (\ref{h0}). Note that $\bar X$ is invariant under three point group operations: (i) 
$x \rightarrow -x$ reflection ($M_x$); (ii) $y \rightarrow -y$ reflection ($M_y$); 
(iii) two-fold rotation around surface normal ($C_2$).  
These symmetry operations, plus time reversal transformation $\Theta$, are represented by the following unitary operators in our $k \cdot p$ theory: 
\beq
&M_x&: - i s_x  \nonumber \\
&M_y&: - i \tau_x s_y \nonumber \\
& C_2&: - i \tau_x s_z \nonumber \\
&\Theta&: i s_y K\label{sym}
\eeq
Here $M_x$ preserves the $L_1$ and $L_2$ valley in the bulk and only acts on electron's spin, whereas 
both $M_y$ and $C_2$ interchange $L_1$ and $L_2$ and hence involve a flavor-changing operator $\tau_x$.   
To zero-th order in $\bk$, we find  two symmetry-allowed operators: $\tau_x$ and $\tau_y s_x$. Therefore   
our $k \cdot p$ Hamiltonian for (001) surface states is given by: 
\beq
H_{{\bar X}_1}(\bk) =  (v_x k_x   s_y - v_y k_y s_x ) + m  \tau_x + \delta s_x    \tau_y. \label{001}
\eeq
Note that the two additional terms, parameterized by $m$ and $\delta$,  
are off-diagonal in flavor space, which correctly describe inter-valley scattering at  the lattice scale.

The $k\cdot p$ Hamiltonian (\ref{001}) is a main result of this work. We now show that $H_{\bar{X}_1}(\bk)$ captures all the essential features of type-II surface states.   
By diagonalizing $H_{{\bar X}_1}(\bk)$, we obtain four surface bands with energy-momentum dispersions $E_H(\bk), -E_H(\bk), E_L(\bk)$ and $-E_L(\bk)$ respectively, where $E_{H, L}(\bk)$ is given by
\begin{widetext}
\beq
E_{H, L}(\bk) = \sqrt{ m^2 + \delta^2 + v_x^2 k_x^2 + v_y^2 k_y^2  \pm 2\sqrt{m^2 v_x^2 k_x^2 + (m^2 + \delta^2) k_y^2 v_y^2 } }.  \label{ek}
\eeq 
\end{widetext}
The resulting surface band structure is plotted in Fig.3. We find two high-energy bands $\pm E_H$ which start from energy $E^X \equiv \sqrt{m^2 + \delta^2}$ at $\bar X$ and coexist in energy with bulk bands, as well as two low-energy  bands $\pm E_L$ which mostly lie within the band gap. 

\begin{figure}[tbp]
\includegraphics[width=8cm]{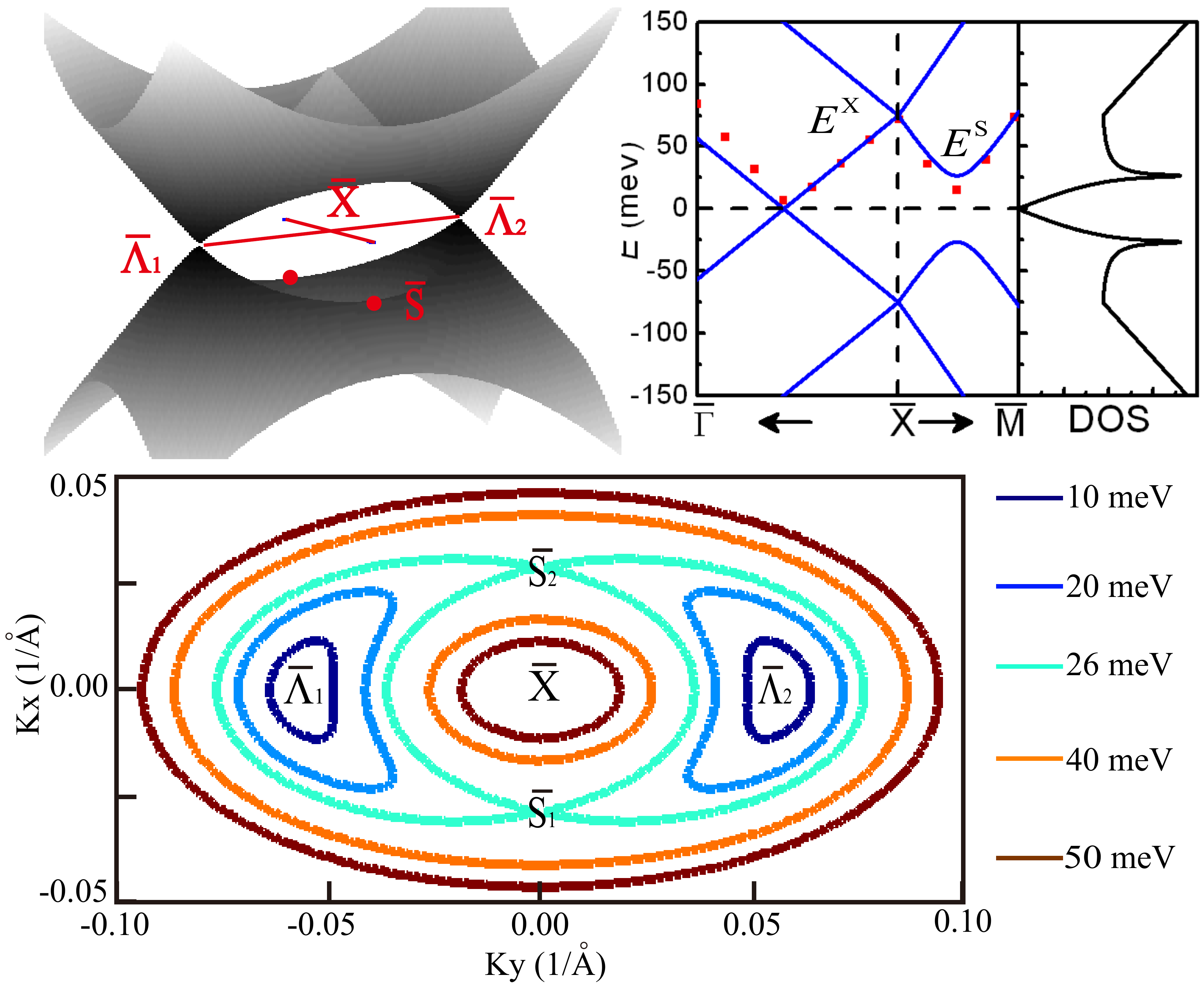}
\caption{$k \cdot p$ band structure for (001) surface states. A pair of low-energy Dirac cones, located at $\bar \Lambda_1$ and $\bar \Lambda_2$ on the line $\bar X \bar \Gamma$, is formed by the interaction between two high-energy Dirac bands centered at $\bar X$. $k \cdot p$ parameters are obtained by fitting with ab-initio results (shown by red dots) on SnTe taken from Ref.\cite{hsieh}: 
$v_x=2.4$eV$\cdot$$\textrm{\AA}$, $v_y=1.3$eV$\cdot$$\textrm{\AA}$, $m=70$meV, and $\delta=26$meV. 
Constant-energy contour evolves rapidly with increasing energy from the Dirac point, changing from two disconnected electron pockets to a large electron pocket and a small hole pocket via a Lifshitz transition. At this transition point,  a saddle point $\bar S$ 
on the line $\bar X \bar M$ leads to a Van-Hove singularity in density of states at energy $E^S = \delta$. }
\end{figure}

It is clear from (\ref{001}, \ref{ek}) that the two terms $m$ and $\delta$ arising from the lattice scale play a key role in shaping the band structure in Fig.3. 
First, a finite $m$ turns the two flavors of massless Dirac fermions into a ``bonding'' ($\tau_x=1$) 
and an ``anti-bonding'' ($\tau_x=-1$) Dirac cones, which are centered at ${\bar X}$ and have energy $\pm m$ respectively.  
In the absence of $\delta$, the lower band 
of the upper Dirac cone and the upper band  of the lower Dirac cone  cross each other at $E=0$ over an elliptical 
contour $C$ in momentum space 
defined by $v_x^2 k_x^2 + v_y^2 k_y^2 = m^2$. Next, a small $\delta$  
turns this band crossing into an anti-crossing via hybridization. 
Importantly, the hybridization matrix element depends on the direction of $\bf k$, and  
leads to a $p$-wave hybridization gap over the contour:   
$
\Delta(\bk) =2 \delta \cdot  v_x k_x /m, \; \bk \in C.     
$
Importantly, $\Delta(\bk)$ vanishes along the mirror-symmetric line $\bar \Gamma \bar X_1$ corresponding to $k_x=0$. 
This is a consequence of the unique electronic topology of the TCI protected by mirror symmetry.   
As can be seen from (\ref{sym}) and (\ref{001}), the two low-energy bands $\pm E_L$ have 
{\it opposite} $M_x$ mirror eigenvalues  on the $k_y$ line $ {\bar X}_1 \bar \Gamma$, 
but  {\it identical} $M_y$ mirror eigenvalues on the $k_x$ line $ {\bar X}_1 \bar M$. 
As a result, hybridization is strictly forbidden on $ {\bar X}_1 \bar \Gamma$, but allowed on ${\bar X}_1 \bar M$. 
The presence of such a protected band crossing on $ \bar X_1 \bar \Gamma$, but not elsewhere, leads to a pair of zero-energy Dirac points $\bar \Lambda_{1,2}$ located  symmetrically away from ${\bar X}_1$ at momenta
$\bar \Lambda_{1,2} = (0, \pm \sqrt{m^2+\delta^2}/v_y)$. By linearizing band structure near each $\bar \Lambda$, we obtain the two-component massless Dirac fermion of Ref.\cite{hsieh} 
\beq
H_{\bar \Lambda}(\delta \bk) = \tilde{v}_x \delta k_x  \sigma_y - v_y \delta k_y  \sigma_x 
\eeq
where $\delta \bk  \equiv \bk - \bar \Lambda$ and the Dirac velocity along $\bar \Gamma \bar X_1$ is reduced from $v_x$:   
$\tilde{v}_x =v_x \delta/\sqrt{m^2 + \delta^2}$.

Our $k\cdot p$ theory  also  
captures essential features of (001) surface states at higher energy, found in the previous ab-initio calculation\cite{hsieh}. 
Fig.3 shows the band dispersion and constant energy contours of the $k \cdot p$ model. 
The surface band structure evolves rapidly with increasing energy away from the Dirac point. 
For $E<\delta$, the Fermi surface consists of two disconnected small Dirac pockets outside $\bar X$. 
As $E$ increases, the two electron pockets first grow in size,  and then touch each other to transform into a large electron pocket and 
a small hole pocket, both centered at $\bar X$. This change of Fermi surface topology (Lifshitz transition) occurs at 
a {\it saddle} point in the band structure located at momentum $\bar S=(m/v_x,0)$ and energy   
$|E^S| = \delta$. The band dispersion near the saddle point is given by
\beq
E(\delta \bk) = E^S + \frac{\delta k_x^2}{2 m_{xx}} - \frac{\delta k_y^2}{2 m_{yy}} , 
\eeq
where  $\delta \bk \equiv \bk - \bar S$ and   
the effective mass tensor is defined by $m_{xx}=\delta / v_x^2$ and $m_{yy}= m^2/( \delta \cdot v_y^2)$. 
This saddle point results in a Van-Hove singularity in density of states, shown in  Fig.3. 

We also compare the band structure from $k\cdot p$ theory (\ref{ek}) with the ab-initio result on SnTe (001) surface\cite{hsieh}, shown by the fitting in Fig.3. The quantitative agreement clearly demonstrates the validity of our $k\cdot p$ theory in a wide energy range measured from the Dirac point. We point out that the fitting can be further improved by further including $\bk$-linear inter-valley terms in $k\cdot p$ theory, in addition to the zero-th terms considered so far (see appendix). This leads to a more sophisticated $k\cdot$ Hamiltonian with seven independent parameters, which is related to a recent study by Fang {\it et al}\cite{kp}. 
We note that these $\bk$-linear terms merely modifies the result of  $H_{\bar X_1}(\bk)$ quantitatively, but do not generate the double-cone band structure on their own. 

\begin{figure}[tbp]
\includegraphics[width=7.5cm]{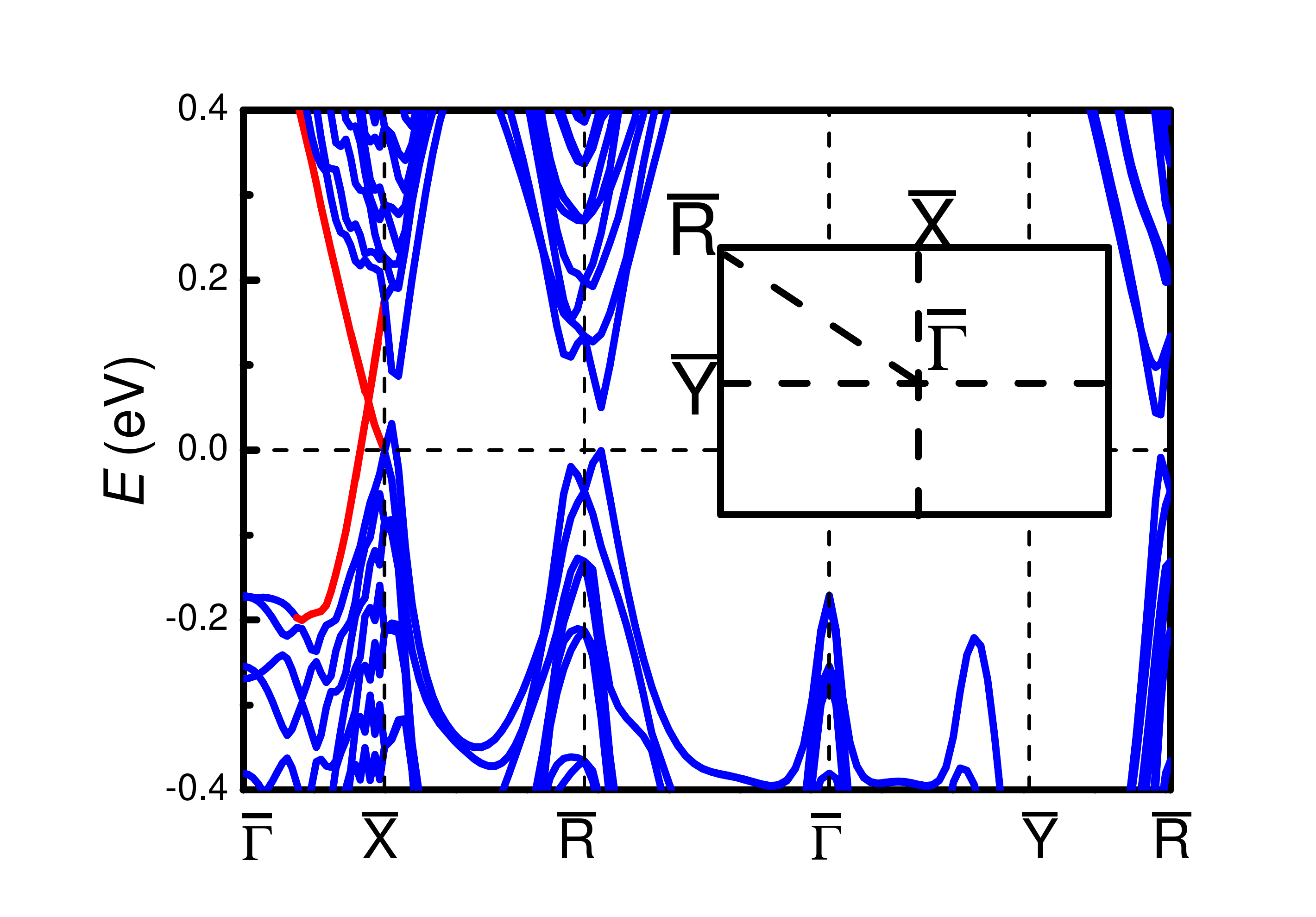}
\caption{Band structure of SnTe (110) surface, from our ab-initio calculations.  The inset shows the surface Brillouin zone.  A pair of Dirac cones is present on the line $\bar X - \bar \Gamma -\bar X$, but absent along other high-symmetry lines.}
\end{figure}

{\bf (110) surface}: we end by briefly discussing another type-II surface, (110).
In this case, $L_1$ and $L_2$ are projected to $\bar X$, and $L_3$ and $L_4$ are projected to $\bar R$. 
Bulk-boundary correspondence based on the electronic  topology of TCI predicts the existence of a pair of counter-propagating states  with opposite mirror eigenvalues on $\bar \Gamma \bar X$\cite{hsieh}. This is confirmed by our ab-initio calculation\cite{method} for SnTe (110) in Fig.4, showing a pair of Dirac cones on the line $\bar X - \bar \Gamma -\bar X$, but not along other high-symmetry lines.  Since $\bar X$ on the (110) surface has the same symmetry as $\bar X_1$ on (001) surface (both have two mirror planes plus a two-fold axis), our $k \cdot p$ theory (\ref{sym}) and (\ref{001}) applies equally well to (110) surface band structure near $\bar X$.  

{\it Note}: Near the completion of this work, we became aware of a preprint on tight-binding calculation of (Pb,Sn)Te surface states\cite{tb}. We also learned of a 
related work by Y. J. Wang {\it et al}.

We thank Timothy Hsieh and Hsin Lin for helpful discussions and earlier collaborations. 
This work is supported by start-up funds from MIT. JL and WD acknowledge the support of the Ministry of  Science and Technology of China (Grant Nos. 2011CB921901 and 2011CB606405), and the National Natural Science Foundation of China  (Grant No. 11074139).

\section{Appendix}

In the main text, we considered lattice-scale terms that are zero-th order in $\bk$ in studying (001) surface states. 
For completeness, we now further include $\bk$-linear terms that are invariant symmetry transformations (\ref{sym}) and examine their effects on the (001) surface band structure. 
We find four such terms: $k_x s_z   \tau_z  $, $k_x  s_y  \tau_x$, $k_y    s_x \tau_x $ and $ k_y \tau_y$. 
This leads to a more sophisticated $k\cdot p$ Hamiltonian for (001) surface: 
\begin{widetext} 
\beq
H'_{{\bar X}_1}(\bk) =  (v_x k_x   s_y - v_y k_y s_x)   + v_x' k_x  s_y  \tau_x + v_x'' k_x   s_z  \tau_z - v_y' k_y  s_x   \tau_x + v_y'' k_y \tau_y + m \tau_x + \delta  s_x \tau_y. \label{h'} 
\eeq
\end{widetext}
We further simplify this Hamiltonian by performing a unitary transformation $U(\theta) \equiv e^{i  \theta s_x \tau_z/2}$. 
Importantly, the form of all symmetry operators (\ref{sym}) is preserved after this transformation. By choosing $\theta$ properly, 
we can eliminate the term $v_x''$ without generating any new term. Therefore the full $k\cdot p$ theory for (001) surface, up to first order in $\bk$, is given by
\begin{widetext} 
\beq
\tilde{H}_{{\bar X}_1}(\bk) =  (v_x k_x   s_y - v_y k_y s_x)   + (v_x' k_x  s_y  - v_y' k_y  s_x)   \tau_x + v_y'' k_y \tau_y + m \tau_x + \delta  s_x \tau_y . \label{full} 
\eeq
\end{widetext}
The first two terms in  (\ref{full}) are inherited from the continuum theory (\ref{h0}) without interaction between valleys. Remarkably,  
all remaining terms  are off-diagonal in valley space, in accordance with their origin from inter-valley mixing at the lattice scale. 
The natural separation between the continuum and the lattice-scale  is the very basis 
of our theory of topological surface states in the TCI phase of IV-VI semiconductors.   

A recent study\cite{kp} obtained a $k\cdot p$ model for (001) surface states, which is equivalent to (\ref{h'}) after a unitary transformation and hence can be further simplified into (\ref{full}) by eliminating a redundant parameter. 
Moreover, the model in Ref.\cite{kp} is constructed {\it entirely} by symmetry analysis, and therefore the physical meaning of the basis states in the model 
is not understood. 

The $\bk$-linear term $(v_x' k_x  s_y  - v_y' k_y  s_x) \tau_x + v_y'' k_y \tau_y$ in the full $k\cdot p$ theory, which is neglected in the main text, plays a secondary role compared to the zero-th order term $m \tau_x + \delta  s_x \tau_y$. Its main effect is to break the particle-hole symmetry of the surface band structure (\ref{001}, \ref{ek}), without changing any of the essential physics. Since narrow-gap IV-VI semiconductors have nearly mirror-like conduction and valence band at low energy, it is a good approximation to neglect this $\bk$-linear term when addressing the main points. We do expect that including this term will further improve the agreement between theory and experiment.

\bibliographystyle{apsrev}

\begin{thebibliography}{10}
\bibitem{method}
We performed density functional theory calculations using the generalized gradient approximation\cite{gga} and the projector augmented wave potential\cite{awp}, as implemented in the Vienna ab initio simulation package\cite{package}.
\end{thebibliography}

\newpage

\end{document}